\documentclass[twocolumn,prd,aps,showpacs,amsmath,amssymb]{revtex4-1}
\usepackage{graphics}
\usepackage{graphicx}
\usepackage{epstopdf}
\usepackage{dcolumn}
\usepackage{bm}
\usepackage{mathrsfs}
\usepackage{pstricks}
\usepackage{color}
\usepackage{slashed}
\usepackage{amsmath}
\usepackage{epsfig}
\usepackage{amsfonts}
\usepackage{amssymb}

\def\beq{\begin{equation}}
\def\eeq{\end{equation}}
\def\bea{\begin{eqnarray}}
\def\eea{\end{eqnarray}}
\def\nn{\nonumber}
\begin{document}

\title{Vacuum fluctuations and radiation reaction in radiative processes of entangled states}
\author{G. Menezes}
\email{gabrielmenezes@ufrrj.br}
\affiliation{Grupo de F\'isica Te\'orica e Matem\'atica F\'isica, Departamento de F\'isica, Universidade Federal Rural do Rio de Janeiro, 23897-000 Serop\'edica, RJ, Brazil}
\author{N. F. Svaiter}
\email{nfuxsvai@cbpf.br}
\affiliation{Centro Brasileiro de Pesquisas F\'{\i}sicas, 22290-180 Rio de Janeiro, RJ, Brazil}

\begin{abstract}
We investigate radiative processes of inertial two-level atoms in an entangled state interacting with a quantum electromagnetic field.  Our intention is to clarify and to analyze the contributions of vacuum fluctuations and radiation reaction to the decay rate of the entangled state. The possible relevance of the findings in the present work is discussed.
\end{abstract}


\maketitle

\section{Introduction}
\label{intro}

Superposition and entanglement are properties that distinguish quantum mechanics from any classical theory. In an entangled quantum system there are states that cannot be factorized into product of states of the subsystems. Entanglement has became of interest since it is a key property in quantum information, cryptography and quantum computation~\cite{1,haroche}. There exist several sources of entangled quantum systems. For instance, in solid state physics, quantum optics and also atoms in cavity quantum electrodynamics. In the literature many ways were proposed to generate entangled states in systems of two-level atoms interacting with a bosonic field. See for example~\cite{2,3,4,5,ved}.

Researches on atomic radiative processes have proved to be of major importance in quantum optics. Following the seminal work of Dicke~\cite{dicke}, there is an extensive literature on resonant interaction of atoms~\cite{mil,aga1,lenz,guo,berman}. Recent examples, of great current interest, concern investigations of radiative processes involving entanglement. Let us briefly discuss some of such important works. For instance, in Ref.~\cite{yang} the authors investigate the properties of emission from two entangled atoms coupled with an electromagnetic field in unbounded space. In Ref.~\cite{eberly} the authors study the radiative processes of entangled two-level atoms coupled individually to two spatially separated cavities. Ref.~\cite{yun} presents an interesting scheme to realize a highly controlled and selective atom-field interaction in cavity quantum electrodynamics system. Nonlinear optical processes in chains of ions in an entangled state were investigated in Ref.~\cite{aga}. See also Ref.~\cite{fi}. We mention that radiative processes of maximally entangled states are quite different from the non-entangled states. For example, in an entangled two-atom system there are superradiant and subradiant states. These states have very different spontaneous decay rates. The symmetric state decays with an enhanced whereas the antisymmetric states decays with a reduced spontaneous emission rate~\cite{rep}.

By using first-order perturbation theory it can be shown that the transition rate of an atom interacting with an electromagnetic field in the vacuum state is given by the Fourier transform of the positive frequency Wightman function evaluated on the world line of the atom~\cite{birrel,nami}. The heuristic picture is then that the atom is forced to radiate by the vacuum field fluctuations. On the other hand, following the discussion by Ackerhalt, Knight and Eberly~\cite{ake} it is possible to interpret spontaneous decay as a radiation-reaction effect. As discussed by Milonni, both effects, vacuum fluctuations and radiation reaction, depend on a particular ordering chosen for commuting atomic and field operators~\cite{mil1}. Hence, it turns out that the contributions of vacuum fluctuations and radiation reaction can to a large extent be chosen arbitrarily. On the other hand, Dalibard, Dupont-Roc, and Cohen-Tannoudji argued that there exists a preferred operator ordering: when one chooses a symmetric ordering the distinct contributions of vacuum fluctuations and radiation reaction to the rate of change of an atomic observable are separately Hermitian and hence they possess an independent physical meaning~\cite{cohen2,cohen3}. This became known as the DDC formalism. Within such a formalism the interplay between vacuum fluctuations and radiation reaction can be considered to maintain the stability of the atom in its ground state. We remark that this formalism was employed in many situations~\cite{aud1,aud2,yu1,rizz,yu2}. The essential picture that emerges from such analysis is the following. The fluctuations of the quantum field act on the atoms, causing their polarizations; the atoms then fluctuate and disturb the field, which in turn reacts back on the atoms. It is clear that the proper comprehension of the interplay between both effects is crucial in the studies of resonant interaction of atoms.

The dynamics of a small system, entangled or not, coupled to a reservoir can be discussed using different formalisms~\cite{scully}. The key point is that the reservoir is generally described by an infinitely many degrees-of-freedom formalism. The standard formalism for the evaluations of time evolution and correlation properties of collective atomic systems is the traditional master equation approach. Within this approach it is possible to quantify the degree of entanglement of a particular state. Here since we are interested in understanding the mechanism responsible for supporting entanglement in radiative processes involving atoms, we choose a different route and use the DDC formalism aforementioned. For an interesting investigation regarding boundary effects on quantum entanglement, see for instance~\cite{hu}.

We choose to go focusing our attention in applying this method to quantum entanglement of atoms. Being more specific, we are interested to study the contributions of vacuum fluctuations and radiation reaction to the stability or decay of an entangled state. We remark that in the present paper we do not consider the rotating-wave approximation. The organization of the paper is as follows. In Section II we discuss the Hamiltonian describing a system of bound atoms and radiation field. In Section III we calculate the rate of variation of the atomic energy in vacuum when both atoms are at rest. Conclusions and final remarks are given in section IV. We briefly review the discussion on relevant electromagnetic correlation functions in the Appendix. In this paper we use units $ \hbar = c = k_B = 1$. We are using the Minkowski metric $\eta_{\alpha\beta} = -1, \alpha=\beta=1,2,3$, $\eta_{\alpha\beta} = 1, \alpha=\beta=0$ and $\eta_{\alpha\beta} = 0,\alpha \neq \beta$.

\section{Two identical atoms coupled with an electromagnetic field}
\label{model}

Let us consider two globally neutral systems of charges localized at fixed distances ${\bf r}_1$ and ${\bf r}_2$ interacting with a quantum electromagnetic field. In what follows we will interpret such systems as identical atoms. We are working in a four-dimensional Minkowski space-time. The Hamiltonian in the Coulomb gauge describing this system in the dipole approximation reads~\cite{cohen}
\bea
H' &=& \sum_{\alpha}\,\frac{1}{2\,m_{\alpha}}\left[{\bf p}_{\alpha} - q_{\alpha}{\bf A}({\bf r}_1)\right]^2
\nn\\
&+&\, \sum_{\beta}\,\frac{1}{2\,m_{\beta}}\left[{\bf p}_{\beta} - q_{\beta}{\bf A}({\bf r}_2)\right]^2
\nn\\
&+&\, V^{11}_{\textrm{Coul}} + V^{22}_{\textrm{Coul}} + V^{12}_{\textrm{dip dip}}
\nn\\
&+&\, \sum_{\lambda=1}^{2}\int \frac{d^3 k}{(2\pi)^3}\, \omega_{{\bf k}}\,a^{\dagger}_{{\bf k},\lambda} a_{{\bf k},\lambda}
\eea
where $V^{11}_{\textrm{Coul}}$ ($V^{22}_{\textrm{Coul}}$) is the respectively Coulomb energy of the system of charges and $V^{12}_{\textrm{dip dip}}$ is the electrostatic interaction energy between the atomic electric dipole moments $\boldsymbol\mu_1$ and $\boldsymbol\mu_2$ of the systems of charges -- the dipole-dipole interaction energy. The sums are over all charges comprising each of the systems. The last part is the Hamiltonian of the free electromagnetic field, in which $a^{\dagger}_{{\bf k},\lambda}, a_{{\bf k},\lambda}$ represent the usual creation and annihilation operators of the electromagnetic field. In the above we have neglected the rest mass energy of the systems of charges and the zero-point energy associated with the quantum field (the energy of the longitudinal field can be considered as a correction to the mechanical rest mass). One expects the dipole approximation to be accurate whenever the coupling between particles and radiation mostly encompasses the modes whose wavelength is much larger than the typical size of the system. The above expression is known as the minimal-coupling form of the Hamiltonian.

The above Hamiltonian can be converted to a more convenient and plain form under the Power-Zienau-Woolley transformation~\cite{pow,woo}. One gets
\bea
H &=& \sum_{\alpha}\,\frac{{\bf p}_{\alpha}^2}{2\,m_{\alpha}} + V^{11}_{\textrm{Coul}}
+ \sum_{\beta}\,\frac{{\bf p}_{\beta}^2}{2\,m_{\beta}} + V^{22}_{\textrm{Coul}}
\nn\\
&+&\, \sum_{\lambda=1}^{2}\int \frac{d^3 k}{(2\pi)^3}\, \omega_{{\bf k}}\,a^{\dagger}_{{\bf k},\lambda} a_{{\bf k},\lambda}
-\boldsymbol\mu_1\cdot{\bf E}(x_1) -\boldsymbol\mu_2\cdot{\bf E}(x_2).
\nn\\
\label{ham-pzw}
\eea
The dipole-dipole interaction energy has been compensated for by dipole terms which also give rise to dipole self-energies. The latter have been neglected in the above expression since it does not contribute to the coupling between the system of charges and the quantum field.  In fact, one can show that the contributions of cross terms related with the transverse polarizations of the systems of charges cancel the contribution coming from $V^{12}_{\textrm{dip dip}}$~\cite{cohen}. The total electric field in the expression above is to be interpreted as an electric displacement. One should understand that the coupling term $\boldsymbol\mu_A\cdot{\bf E}({\bf r}_A)$ contains the interaction of the system of charges $A$ with the transverse electric (displacement) field as well as the longitudinal field generated by the other system of charges at ${\bf r}_A$. The consequence of such considerations is that there are no more instantaneous electrostatic  interaction terms between the systems of charges; all interactions now are realized through the quantum electromagnetic fields which propagate with velocity $1$. This form is suitable for describing retarded dipole-dipole interactions between the atoms. Expression~(\ref{ham-pzw}) is known as the multipolar-coupling form of the Hamiltonian. This is the representation that we shall consider throughout this paper.

Hereafter we will be working within the Heisenberg picture. As mentioned above, we assume that the globally neutral systems of charges describe two identical two-level atoms at rest located at ${\bf r}_1$ and ${\bf r}_2$, respectively. We denote by
$|g_i\rangle$ and $|e_i\rangle$ the ground and excited state of the $i$-th atom, with energies $\mp\,\omega_0/2$, respectively. We employ this notation in order to allow the possibility of a straighforward generalization of our results to the case of non-identical atoms. The time evolution of the total system is to be taken with respect to the proper time $\tau$ of the atoms. The important fact to be noticed is that the purely atomic parts that appear in equation~(\ref{ham-pzw}) describe the free Hamiltonian of the two atoms labelled as $H_A(\tau)$. In the Dicke notation~\cite{dicke}, one has
\beq
H_A(\tau) = \frac{\omega_0}{2}\left[\left(S_{1}^z(\tau)\otimes\hat{1}\right)\,+\left(\hat{1}\otimes S_{2}^z(\tau)\right)\right],
\label{ha}
\eeq
where $S_a^z = | e_a \rangle\langle e_a | - | g_a \rangle\langle g_a |$ is the energy operator of the $a$-th atom. The space of the two-atom system is spanned by four product stationary states which are eigenstates of $H_A$ with respective energies
\bea
&& E_{gg} = -\omega_0\,\,\,\,|gg\rangle = |g_1\rangle|g_2\rangle,
\nn\\
&& E_{ge}= 0\,\,\,\,|ge\rangle = |g_1\rangle|e_2\rangle,
\nn\\
&&E_{eg} = 0\,\,\,\,|eg\rangle = |e_1\rangle|g_2\rangle,
\nn\\
&&E_{ee} = \omega_0\,\,\,\,|ee\rangle = |e_1\rangle|e_2\rangle,
\label{sta}
\eea
where a tensor product is implicit. Instead of working with this product-state basis, we can conveniently choose the Bell state basis. In terms of the product states, one has:
\bea
|\Omega^{\pm}\rangle &=& \frac{1}{\sqrt{2}}\left(|g_1\rangle|e_2\rangle \pm |e_1\rangle|g_2\rangle\right)
\nn\\
|\Phi^{\pm}\rangle &=& \frac{1}{\sqrt{2}}\left(|g_1\rangle|g_2\rangle \pm |e_1\rangle|e_2\rangle\right).
\label{bell}
\eea
The Bell states are known as the four maximally entangled two-qubit Bell states, and they form a convenient basis of the two-qubit space. Nevertheless, we remark that the Bell states $|\Phi^{\pm}\rangle$ are not eigenstates of the atomic Hamiltonian $H_A$; in turn, in view of the degeneracy associated with the eigenstates $|ge\rangle$ and $|eg\rangle$, any linear combination of these degenerate eigenstates is also an eigenstate of the atomic Hamiltonian corresponding to the same energy eigenvalue. Therefore, the Bell states $|\Omega^{\pm}\rangle$ are eigenstates of $H_A$.

The stationary trajectories guarantees that the undisturbed atomic system has stationary states. In the following we will be interested in the evolution of the quantum electromagnetic field with respect to the proper time $\tau$. In this way, the Heisenberg equations of motions tells us that
\beq
H_F(\tau) = \sum_{\lambda}\int \frac{d^3 k}{(2\pi)^3}\, \omega_{{\bf k}}\,a^{\dagger}_{{\bf k},\lambda}(t(\tau))a_{{\bf k},\lambda}(t(\tau))\,\frac{dt}{d\tau},
\label{hf}
\eeq
Finally, from equation~(\ref{ham-pzw}) one can easily identify the Hamiltonian which describes the interaction between the atoms and the field in the dipole approximation:
\beq
H_{I}(\tau) = - \boldsymbol\mu_1(\tau) \cdot {\bf E}(x_1(\tau)) - \boldsymbol\mu_2(\tau) \cdot {\bf E}(x_2(\tau)).
\label{hi}
\eeq
The dipole moment operator is endowed of only off-diagonal elements and hence can be written as
\beq
\boldsymbol\mu_i(\tau) = \langle g_i |\boldsymbol\mu_i| e_i \rangle | g_i \rangle\langle e_i| + \textrm{H.c.} \equiv \boldsymbol\mu\left[S_{i}^{+}(\tau) + S_{i}^{-}(\tau)\right],
\label{dip}
\eeq
(no summation over repeated indices) where we have assumed a proper choice of phases which allows the dipole matrix elements to be real and we have defined the dipole raising and lowering operators as $S_{i}^{+} = | e_i \rangle\langle g_i|$ and $S_{i}^{-} = | g_i \rangle\langle e_i|$, respectively. In addition, since the atoms are presumed to be identical and similarly oriented, $\langle g_i |\boldsymbol\mu_i| e_i \rangle$ is independent of the index $i$ and denoted simply by $\boldsymbol\mu$. Incidentally, suppose that our atoms are spinless one-electron systems. Hence $\boldsymbol\mu_a(\tau) = e\,{\bf r}_a(\tau)$, where $e$ is the electron charge and ${\bf r}_a(\tau)$ is the position operator of the atom $a$.

As discussed above, there is an extensive literature investigating the role of vacuum fluctuations and radiation reaction in radiative processes of atoms. Here our main intention is to identify the distinct contributions of quantum field vacuum fluctuations and radiation reaction to the entanglement dynamics of the atoms. For such purposes, let us briefly discuss the DDC formalism. The first step is to set up the the Heisenberg equations of motion for the dynamical variables of the atom and the field with respect to $\tau$. Such expressions can be derived from the total Hamiltonian $ H(\tau) = H_A(\tau) + H_F(\tau) + H_I(\tau)$. Afterwards one splits the solutions of the equations of motion in two parts, namely: The free part, which is present even in the absence of the coupling; and the source part, which is caused by the interaction between atoms and field. Therefore one can write, for the atomic observables and the dynamical variables of the field, respectively:
\bea
S_a^z(\tau) &=& S_a^{z,f}(\tau) + S_a^{z,s}(\tau),
\nn\\
a_{{\bf k},\lambda}(t(\tau)) &=& a^{f}_{{\bf k},\lambda}(t(\tau)) + a^{s}_{{\bf k},\lambda}(t(\tau)).
\eea
$a = 1, 2$. Since one can construct from the annihilation and creation field operators the free and source part of the quantum electric field, one also has
\beq
{\bf E}(t(\tau),{\bf x}(\tau)) = {\bf E}^{f}(t(\tau),{\bf x}(\tau)) + {\bf E}^{s}(t(\tau),{\bf x}(\tau)).
\eeq
Nevertheless, such a procedure introduces an ambiguity of operator ordering. The source part of the field acquires contributions of atomic observables during its time evolution which implies that the fundamental aspect that all atomic observables commute with field variables is not preserved in time for ${\bf E}^{f}$ and ${\bf E}^s$ separately.  Because the above expression contains products of atomic and field operators, one must choose an operator ordering when discussing the effects of ${\bf E}^{f}$ and ${\bf E}^s$ separately. Different operator orderings will yield different interpretations regarding the roles played by vacuum fluctuations and radiation reaction. As pointed out in the Introduction, only the adoption of a symmetric ordering of atomic and field variables permits both contributions to be Hermitian. In this way, the effects of such phenomena can posses an independent physical meaning. Therefore, following this prescription, one can clearly separate the contribution of the vacuum fluctuations from the radiation-reaction effects in the evolution of the atoms' energies, which are given by the expectation value of $H_A$, which in turn is given by equation~(\ref{ha}). Furthermore, in a perturbative treatment, we take into account only terms up to order $e^2$. We also consider an averaging over the field degrees of freedom by taking vacuum expectation values. In turn, since we are interested in the evolution of expectation values of atomic observables, we also take the expectation value of the above expressions in a state $|\omega\rangle$, which can be one of the states given by equation~(\ref{sta}) or one can also consider the Bell states~(\ref{bell}). The details of the calculations can be found elsewhere, see for instance~\cite{aud1}. Therefore, employing the notation $\langle (\cdots) \rangle = \langle 0,\omega |(\cdots)| 0,\omega \rangle$, one has the pivotal results concerning the vacuum-fluctuation contribution
\bea
\Biggl\langle \frac{d H_A}{d\tau} \Biggr\rangle_{VF} &=& \frac{i}{2}\int_{\tau_0}^{\tau}d\tau' \,\sum_{a,b = 1}^{2}D_{ij}(x_a(\tau),x_b(\tau'))
\nn\\
&\times&\,\frac{d}{d\tau}\Delta^{ij}_{ab}(\tau,\tau'),
\label{vfha3}
\eea
where
\beq
\Delta^{ij}_{ab}(\tau,\tau') = \langle \omega | [\mu^{i,f}_{a}(\tau),\mu^{j,f}_{b}(\tau')]| \omega \rangle,\,\,\,a,b = 1,2,
\label{susa}
\eeq
is the linear susceptibility of the two-atom system in the state $|\omega\rangle$ and
\beq
D_{ij}(x_a(\tau),x_b(\tau')) = \langle 0 |\{E^{f}_{i}(x_a(\tau)),E^{f}_{j}(x_b(\tau'))\}| 0 \rangle,
\eeq
is the Hadamard's elementary function. Since we are dealing with free fields, one can employ the results derived in the Appendix~\ref{A}. On the other hand, for the radiation-reaction contribution, one has:
\bea
\Biggl\langle \frac{d H_A}{d\tau} \Biggr\rangle_{RR} &=& \frac{i}{2}\int_{\tau_0}^{\tau}d\tau' \,\sum_{a,b=1}^{2}\Delta_{ij}(x_a(\tau),x_b(\tau'))
\nn\\
&\times&\,\frac{d}{d\tau}D^{ij}_{ab}(\tau,\tau'),
\label{rrha3}
\eea
where
\beq
D^{ij}_{ab}(\tau,\tau') = \langle \omega | \{\mu^{i,f}_{a}(\tau),\mu^{j,f}_{b}(\tau')\}| \omega \rangle,\,\,\,a,b = 1,2,
\label{cora}
\eeq
is the symmetric correlation function of the two-atom system in the state $|\omega\rangle$ and
\beq
\Delta_{ij}(x_a(\tau),x_b(\tau')) = \langle 0 |[E^{f}_{i}(x_a(\tau)),E^{f}_{j}(x_b(\tau'))]| 0 \rangle,
\eeq
is the Pauli-Jordan function which, for the same reason as above, can be evaluated from Appendix~\ref{A}. Note that $\Delta^{ij}_{ab}$ and $D^{ij}_{ab}$ characterize only the two-atom system itself, unlike the statistical functions of the field which have to be evaluated along the trajectory of the atoms. We see from equations~(\ref{vfha3}) and~(\ref{rrha3}) that the rate of variation of the energy of the two-atom system presents contributions from the isolated atoms and also contributions due to cross correlations between the atoms mediated by the field. This interference is a consequence of the interaction of each atom with the field.

Using that $\mu^{i,f}_{a}(\tau) = e^{i H_A\tau}\mu^{i,f}_{a}(0)e^{-i H_A\tau}$, and the completeness relation for the two-atom states $\sum_{\omega'}| \omega' \rangle \langle \omega' | = {\bf 1}$, one can show that the statistical functions for the two-atom system can be put in the following explicit forms
\bea
&&\Delta^{ij}_{ab}(\tau,\tau') = \sum_{\omega'} \biggl[{\cal M}^{ij}_{ab}(\omega,\omega')\,e^{i(\omega - \omega')(\tau - \tau')}
\nn\\
&&-{\cal M}^{ji}_{ba}(\omega,\omega')\,e^{-i(\omega - \omega')(\tau - \tau')}\biggr],
\label{susa1}
\eea
and
\bea
&&D^{ij}_{ab}(\tau,\tau') = \sum_{\omega'} \biggl[{\cal M}^{ij}_{ab}(\omega,\omega')\,e^{i(\omega - \omega')(\tau - \tau')}
\nn\\
&&+ {\cal M}^{ji}_{ba}(\omega,\omega')\,e^{-i(\omega - \omega')(\tau - \tau')}\biggr],
\label{cora1}
\eea
where we have defined
\beq
{\cal M}^{ij}_{ab}(\omega,\omega') = \langle \omega |\mu^{i,f}_{a}(0)| \omega' \rangle\langle \omega' |\mu^{j,f}_{b}(0)| \omega \rangle.
\label{aa}
\eeq
In this paper we are primarily interested in transitions from entangled states [currently represented by the Bell states~(\ref{bell})] to separable states [given by equation~(\ref{sta})], or the inverse. For the purpose of investigating the fine details regarding the degradation of entanglement between the atoms as a spontaneous emission phenomenon, assume that the atoms were initially prepared in one of the Bell states $|\Omega^{\pm}\rangle$. With this regard, equations~(\ref{sta}) and~(\ref{bell}) state that the only allowed transitions are $|\Omega^{\pm}\rangle \to |gg\rangle$, with $\Delta \omega = \omega - \omega' = \omega_0 > 0$ and $|\Omega^{\pm}\rangle \to |ee\rangle$, with $\Delta \omega = \omega - \omega' = - \omega_0 < 0$. On the other hand, suppose that one is interested to address the generation of entanglement. Assume that the atoms were initially prepared in the excited state $|ee\rangle$. For the Bell states $|\Phi^{\pm}\rangle$ one gets $\Delta \omega = 0$, which implies that $\langle d H_A/dt \rangle = 0$. Hence it is not possible to generate such Bell states out of the excited atoms. One then must consider the decay rate to one of the Bell states $|\Omega^{\pm}\rangle$, obtaining 
$\Delta \omega = \omega_0 > 0$.

\section{Rate of variation of the atomic energy in vacuum for static atoms}
\label{atom}

In this Section we will consider in detail the rate of variation of atomic energy for an inertial two-atom system using the technique developed above. We assume that the atoms are separated by the space-time interval $\Delta x^{\mu} = (x(\tau_1) - x(\tau_2))^{\mu} = (\tau_1 - \tau_2, {\bf x}_{0_1} - {\bf x}_{0_2}) = (\tau_1 - \tau_2, \Delta{\bf x}_0)$. The contributions~(\ref{vfha3}) and~(\ref{rrha3}) to spontaneous emission can be evaluated from the results discussed in the Appendix~\ref{A}. Being more specific, since $x = x(\tau)$ and $x' = x(\tau')$ we insert in such expressions the statistical functions of the two atom system, given by equations~(\ref{susa1}) and~(\ref{cora1}), and the electromagnetic-field statistical functions given by~(\ref{pauli}) and~(\ref{hada}). Initially let us present the contributions coming from the vacuum fluctuations. These are given below, with $u = \tau - \tau'$:
\begin{widetext}
\bea
\Biggl\langle \frac{d H_A}{d\tau} \Biggr\rangle_{VF} &=& \frac{1}{2\pi^2}\sum_{\omega'}\sum_{a,b = 1}^{2}{\cal M}^{ij}_{ab}(\omega,\omega')\Delta\omega \int_{-\Delta\tau}^{\Delta\tau} du\,e^{i\Delta\omega u}
\biggl\{\frac{2(\Delta{\bf x})_i(\Delta{\bf x})_j - \delta_{ij}\left[|\Delta{\bf x}|^2 + (u - i\epsilon)^2\right]}{\left[(u - i\epsilon)^2 - |\Delta{\bf x}|^2\right]^3}
\nn\\
&&+\,\frac{2(\Delta{\bf x})_i(\Delta{\bf x})_j - \delta_{ij}\left[|\Delta{\bf x}|^2 + (u + i\epsilon)^2\right]}{\left[(u + i\epsilon)^2 - |\Delta{\bf x}|^2\right]^3}\biggr\},
\nn\\
\eea
where $\Delta\tau= \tau - \tau_0$, $\Delta{\bf x} = {\bf x}_{0_a} - {\bf x}_{0_b}$.
By invoking the method of residues one may compute all the relevant integrals. For $\epsilon\to 0$ one gets
\bea
\Biggl\langle \frac{d H_A}{d\tau} \Biggr\rangle_{VF} &=& -\frac{1}{4\pi}\sum_{\omega'}\sum_{a,b = 1}^{2}\frac{\Delta\omega\sin(\Delta\omega |\Delta{\bf x}|)}{|\Delta{\bf x}|}
\,{\cal M}^{ij}_{ab}(\omega,\omega')
\nn\\
&\times&\,\Bigl(\theta(\Delta\omega)-\theta(-\Delta\omega)\Bigr)\Biggl[\left(\frac{3(\Delta{\bf x})_i(\Delta{\bf x})_j}{|\Delta{\bf x}|^2} - \delta_{ij}\right)\Biggl(1 - \Delta\omega
|\Delta{\bf x}| \cot(\Delta\omega |\Delta{\bf x}|)\Biggr)\frac{1}{|\Delta{\bf x}|^2}
\nn\\
&&-\left(\frac{(\Delta{\bf x})_i(\Delta{\bf x})_j}{|\Delta{\bf x}|^2} - \delta_{ij}\right)|\Delta\omega|^2\Biggr],
\eea
where we have taken $\Delta \tau \to \infty$. Let us note from the above contributions that vacuum fluctuations stimulate excitation ($\langle d H_A/dt \rangle > 0$) as well as deexcitation ($\langle d H_A/dt \rangle < 0$) of the two-atom system. This is reminiscent from the fact that for an atom interacting with quantized radiation, one has that stimulated excitation and deexcitation have equal Einstein $B$ coeffcients, which implies that vacuum fluctuations tend to excite an atom in the ground state as well as deexcitate an atom in the excited state. In our context, following the discussion at the end of Section~\ref{model} one realizes that vacuum fluctuations can promote the degradation of entanglement of atoms as well as activate the generation of entanglement between the atoms.

Now let us evaluate the radiation-reaction contributions. One has, with $u = \tau - \tau'$:
\bea
\Biggl\langle \frac{d H_A}{d\tau} \Biggr\rangle_{RR} &=& \frac{1}{4\pi}\sum_{\omega'}\sum_{a,b = 1}^{2}{\cal M}^{ij}_{ab}(\omega,\omega')\Delta\omega\int_{0}^{\Delta\tau}du \,\sin(\Delta\omega u)
\nn\\
&&\times\,\Biggl\{\left(\frac{3(\Delta{\bf x})_i(\Delta{\bf x})_j}{|\Delta{\bf x}|^2} - \delta_{ij}\right)\Biggl[\frac{\delta'(|\Delta{\bf x}| - u) - \delta'(|\Delta{\bf x}| + u)}{|\Delta{\bf x}|^2}
-\left(\frac{\delta(|\Delta{\bf x}| - u) - \delta(|\Delta{\bf x}| + u)}{|\Delta{\bf x}|^3}\right)\Biggr]
\nn\\
&&-\left(\frac{(\Delta{\bf x})_i(\Delta{\bf x})_j}{|\Delta{\bf x}|^2} - \delta_{ij}\right)\left[\frac{\delta''(|\Delta{\bf x}| - u) - \delta''(|\Delta{\bf x}| + u)}{|\Delta{\bf x}|}\right]\Biggr\}.
\eea
By employing standard Dirac delta-function identities we get:
\bea
\Biggl\langle \frac{d H_A}{d\tau} \Biggr\rangle_{RR} &=& -\frac{1}{4\pi}\sum_{\omega'}\sum_{a,b = 1}^{2}\frac{\Delta\omega\sin(\Delta\omega |\Delta{\bf x}|)}{|\Delta{\bf x}|}
\Bigl(\theta(\Delta\omega)+\theta(-\Delta\omega)\Bigr){\cal M}^{ij}_{ab}(\omega,\omega')
\nn\\
&&\times\,\Biggl[\left(\frac{3(\Delta{\bf x})_i(\Delta{\bf x})_j}{|\Delta{\bf x}|^2} - \delta_{ij}\right)\Bigl(1 - \Delta\omega|\Delta{\bf x}| \cot(\Delta\omega |\Delta{\bf x}|)\Bigr)\frac{1}{|\Delta{\bf x}|^2}
\nn\\
&&-\left(\frac{(\Delta{\bf x})_i(\Delta{\bf x})_j}{|\Delta{\bf x}|^2} - \delta_{ij}\right)(\Delta\omega)^2\Biggr].
\eea
\end{widetext}
Note that our results explicitly show contributions from a classical phenomenon (radiation-reaction effect) to the entanglement dynamics of two identical atoms. Adding up the above terms leads to the main conclusions of this work. First of all, we note the appearance of spatial oscillations in the rate of variation of atomic energy. In turn, we observe that for large times the difference between vacuum fluctuations and radiation reaction arises when considering excitation processes. In the perspective of spontaneous emission, both give identical contributions. Thus only emission processes are allowed for asymptotic times. Being more specific, the Bell states $|\Omega^{\pm}\rangle$ are stable for absorption processes but through spontaneous emission the atoms can disentangle and decay to the ground state. On the other hand, the transition $|gg\rangle \to |\Omega^{\pm}\rangle$ is suppressed for $\Delta\tau \to \infty$. This can be seen as a direct consequence of the known result that an atom in its ground state is stable due to the balance between vacuum fluctuations and radiation reaction. In order to perturb such a balance for $\Delta\tau \to \infty$ one must consider general non-inertial trajectories, such as uniformly accelerated motion~\cite{aud1}. Notwithstanding one can notice that entanglement between the atoms can still be created, but only via emission processes; for instance, one must initially prepare the atoms in the excited state $|ee\rangle$ in order to detect the formation of entangled states for $\Delta\tau \to \infty$. In summary, the importance of such results lies in the fact that by considering the cooperation between vacuum fluctuations and radiation-reaction effect, one could give a detailed mechanism that accounts for the generation as well as the degradation of entanglement in the resonant interaction of atoms.

\begin{figure}[!htb]
\begin{center}
\includegraphics[scale=.94]{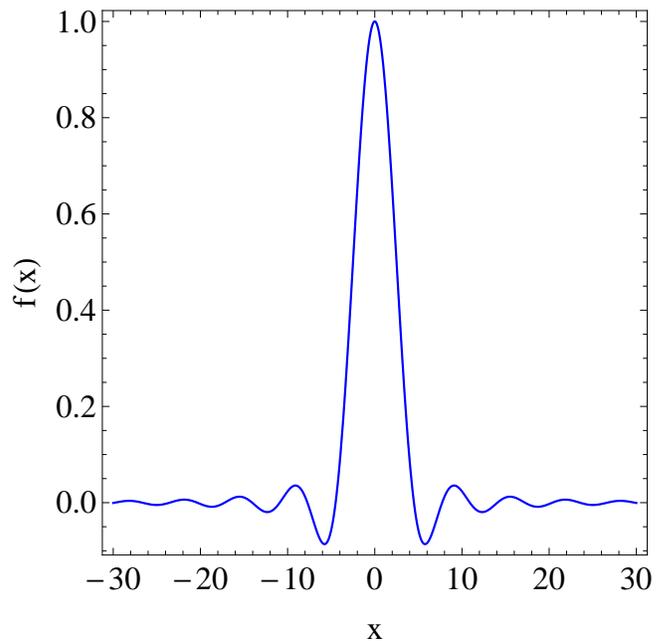}
\caption{The quantity $f(x)$ as a function of $x = \omega_0\Delta z$.}
\label{1}
\end{center}
\end{figure}

For completeness let us present the total rate of change of the atomic energy for $\Delta\tau \to \infty$ assuming that the atoms were initially prepared in one of the Bell states $|\Omega^{\pm}\rangle$. This is obtained by adding the vacuum-fluctuations and radiation-reaction terms. For simplicity consider that the atoms are polarized along the $x_3$ direction. We also assume the special case in which the atoms are separated by the space-time interval $\Delta x^{\mu} = (x(\tau_1) - x(\tau_2))^{\mu} = (\tau_1 - \tau_2, 0, 0, ({\bf x}_{1})_3 - ({\bf x}_{2})_3)$. The total rate of change of the atomic energy is zero for
$\Delta\omega  < 0$. As discussed above the distinct contributions for spontaneous excitation cancel each other in this situation, leaving only the terms for which $\Delta \omega > 0$. In addition, from equation~(\ref{sta}) and the aforementioned assumption on the dipole matrix elements one has that
\bea
{\cal M}^{ij}_{11}(\Omega^{\pm}, g) &=& \frac{\mu^{i}\mu^{j}}{2}
\nn\\
{\cal M}^{ij}_{22}(\Omega^{\pm}, g) &=& \frac{\mu^{i}\mu^{j}}{2}
\nn\\
{\cal M}^{ij}_{12}(\Omega^{\pm}, g) &=& {\cal M}_{21}(\Omega^{\pm}, g) = \pm\, \frac{\mu^{i}\mu^{j}}{2}.
\label{mel}
\eea
Hence the explicit result for the rate of variation of energy of the two-atom system is given by:
\beq
\Biggl\langle \frac{d H_A}{d\tau} \Biggr\rangle_{tot} = -\frac{\omega_0^4}{3\pi}\,\Bigl[1 \pm\, f(\omega_0\Delta z)\Bigr] (\mu^{3})^2,
\label{tot}
\eeq
where the (minus) plus sign refers to the (anti)symmetric Bell state ($|\Omega^{-}\rangle$) $|\Omega^{+}\rangle$ and $\Delta z = ({\bf x}_{1})_3 - ({\bf x}_{2})_3$. Also, we have defined the function:
\beq
f(x) = \frac{3\sin x}{x^3}\,\Bigl(1 - x \cot x\Bigr).
\eeq
Such a function quantifies the influence of the cross correlations on the entanglement between the atoms. Some special values are given by
\beq
f(n \pi) =  \frac{3}{(n \pi)^2}\,(-1)^{n+1},
\eeq
($n$ is a positive integer) and
\beq
f((n + 1/2) \pi) = \frac{3}{(n + 1/2)^3\pi^3}\,(-1)^{n}.
\eeq
The behavior of this function is depicted in figure~\ref{1}. Note the great oscillatory regime for short distances between the atoms. In fact, the plot shows that for $|\Delta z| \ll 1/\omega_0$ cross correlations are more important for the rate $dH_A/d\tau$ in comparison with the case in which the distance between the atoms is very large ($|\Delta z| \gg 1/\omega_0$). It means that the cross correlations generate a constructive interference when the atoms are very near each other and these interference terms vanish for large spatial separations between entangled atoms. This is a similar result as described in references~\cite{lenz,juan}. On the other hand, equation~(\ref{tot}) shows that, for small distances between the atoms, the symmetric Bell state $|\Omega^{+}\rangle$ decays with an enhanced emission (superradiant) rate as compared with terms of isolated atoms, whereas the antisymmetric Bell state $|\Omega^{-}\rangle$ presents the opposite behavior (subradiant). This clearly indicates that the antisymmetric entangled state decays in a softened rate. In addition, note that for $|\Delta z| \ll 1/\omega_0$ the antisymmetric Bell state presents a nearly complete inhibition of the decay to the ground state due to destructive interference of cross correlations between the atoms. Such results confirm the assertions discussed in the Introduction.
%

\section{Conclusions and Perspectives}

Within first-order perturbation theory the transition rate of an atom interacting with an electromagnetic field in the vacuum state is given by the Fourier transform of the positive frequency Wightman function evaluated on the world line of the atom. Within this picture the atom is forced to radiate by the vacuum field fluctuations. On the other hand, it is possible to interpret spontaneous decay as a radiation-reaction effect. Both effects, vacuum fluctuations and radiation reaction, depend on a particular ordering chosen for commuting atomic and field operators. On the other hand, as discussed above there exists a preferred operator ordering: when one chooses a symmetric ordering the distinct contributions of vacuum fluctuations and radiation reaction to the rate of change of an atomic observable are separately Hermitian and hence they possess an independent physical meaning~\cite{cohen2,cohen3}. Within such a formalism the interplay between vacuum fluctuations and radiation reaction can be considered to maintain the stability of the atom in its ground state. Here in this paper we studied the contributions of vacuum fluctuations and radiation reaction in radiative processes of entangled atoms. We found that, at asymptotic times, with respect to excitation processes such contributions cancel each other whereas both mechanisms are responsible for the decay of the entangled state. One possible way to unbalance the contributions from vacuum fluctuations and radiation reaction in absorption processes is to consider situations of finite observation time intervals. The main conclusion one can draw from such results is that when considering radiative processes of atoms the machinery underlying entanglement can be envisaged as an interplay between classical concepts (radiation-reaction effect) and quantum-mechanical phenomena (vacuum fluctuations). We believe that the results presented in this paper may have an impact in the studies of radiative process of atoms. For instance, recently the subject of van der Waals interaction between atoms was taken up in reference~\cite{don} in which the authors demonstrate that such an interaction presents both temporal and spatial oscillations. In the experimental side, the angular dependence of the resonant dipole-dipole interaction between two individual Rydberg atoms with controlled relative positions was measured~\cite{syl}. A framework in which vacuum fluctuations and radiation-reaction effect have been clearly identify and quantitatively analyzed may contribute to an accurate and deeper understanding of such results.

Conducting plates or cavities can modify the rate of spontaneous emission of excited atoms~\cite{mesh}. Since the presence of boundaries affect the vacuum fluctuations of the quantum field one should expect that the transition rates of atoms are modified in this situation~\cite{ford}. Hence, for entangled atoms, it is interesting to ask how the rate of variation of the atomic energy is modified in a situation where translational invariance is broken. Recently, boundary effects on entanglement were investigated assuming a framework where the spontaneous decay of atoms is only attributed to vacuum-fluctuation effects~\cite{juan}. A natural extension of our results is to discuss the mean life of an entangled state in the presence of boundaries within the formalism employed in this work. This is under investigation by the authors.

\section*{Acknowlegements}

This paper was partially supported by Conselho Nacional de Desenvolvimento Cientifico e Tecnol{\'o}gico do Brasil (CNPq).

\appendix

\section{Correlation functions of the electromagnetic fields in the vacuum}
\label{A}

In this appendix we shall concisely discuss free electromagnetic correlation functions. We consider a four dimensional unbounded Minkowski space-time. In addition, as previously mentioned we will work in the Coulomb gauge. Here we will not discuss the quantization of the electromagnetic field in detail. For a thorough analysis of quantum electrodynamics in the Coulomb gauge, we refer the reader the reference~\cite{cohen}. In turn, details concerning the correlations of the electric field for two static space-time points can be found in the comprehensive account by Takagi~\cite{takagi} and related references therein. Instead, we limit ourselves to present all relevant electromagnetic correlation functions in the vacuum employed in the text.

Let us start with the correlation functions of the electric field. Through standard procedures, one may evaluate the Pauli-Jordan function of the electric field. It is given by
\bea
&&\Delta^{ij}(x,x') = \langle 0 |[E^{i}(x), E^{j}(x')]| 0\rangle
\nn\\
&&\,= -i\left(\frac{\partial}{\partial t}\frac{\partial}{\partial t'}\delta^{ij} - \frac{\partial}{\partial x_i}\frac{\partial}{\partial x_j'}\right)\,D(t - t',{\bf x} - {\bf x}'),
\eea
where 
\bea
&&D(t,{\bf x}) = \frac{1}{4\pi|{\bf x}|}\Bigl[\delta(|{\bf x}| - t) - \delta(|{\bf x}| + t)\Bigr]
\nn\\
&& = \frac{1}{(2\pi)^2|{\bf x}|}\lim_{\epsilon \to 0}\left[\frac{\epsilon}{\epsilon^2+(|{\bf x}| - t)^2} - \frac{\epsilon}{\epsilon^2+(|{\bf x}| + t)^2}\right].
\label{pj}
\eea
By evaluating the derivatives one obtains an explicit form for the Pauli-Jordan function in terms of the delta functions and their derivatives:	
\begin{widetext}
\bea
\Delta^{ij}(x,x') &=& \frac{i}{4\pi}\Biggl\{\left(\frac{3(\Delta{\bf x})^i(\Delta{\bf x})^j}{|\Delta{\bf x}|^2} - \delta^{ij}\right)\Biggl[\frac{\delta'(|\Delta{\bf x}| - \Delta t) - \delta'(|\Delta{\bf x}| + \Delta t)}{|\Delta{\bf x}|^2}
\nn\\
&-& \left(\frac{\delta(|\Delta{\bf x}| - \Delta t) - \delta(|\Delta{\bf x}| + \Delta t)}{|\Delta{\bf x}|^3}\right)\Biggr]
\nn\\
&-&\left(\frac{(\Delta{\bf x})^i(\Delta{\bf x})^j}{|\Delta{\bf x}|^2} - \delta^{ij}\right)\left[\frac{\delta''(|\Delta{\bf x}| - \Delta t) - \delta''(|\Delta{\bf x}| + \Delta t)}{|\Delta{\bf x}|}\right]\Biggr\},
\label{pauli}
\eea
\end{widetext}
where $\Delta{\bf x} = {\bf x} - {\bf x}'$ and $\Delta t = t - t'$. The above derivatives of the delta function are to be understood as distributional derivatives. An important remark is the following. The expression we have obtained  is ambiguous when $\Delta t \to 0$ or $|\Delta {\bf x}| \to 0$ since they introduce products of delta functions with functions which diverge for $|\Delta {\bf x}|  = 0$. In this case, one possible route is to consider the second line of expression~(\ref{pj}) with a finite $\epsilon$.

Another important commutator of the electric field is the Hadamard's elementary function. It is given by
\bea
&& D^{ij}(x,x') = \langle 0 |\{E^{i}(x), E^{j}(x')\}| 0\rangle
\nn\\
&&= \left(\frac{\partial}{\partial t}\frac{\partial}{\partial t'}\delta^{ij} - \frac{\partial}{\partial x_i}\frac{\partial}{\partial x_j'}\right)\,D^{(1)}(t-t',{\bf x} - {\bf x}'),
\eea
where
\bea
D^{(1)}(t - t',{\bf x} - {\bf x}') &=& -\frac{1}{4\pi^2}\biggl\{\frac{1}{\left[(t - t' - i\epsilon)^2 - |{\bf x} - {\bf x}'|^2\right]}
\nn\\
&+&\, \frac{1}{\left[(t - t' + i\epsilon)^2 - |{\bf x} - {\bf x}'|^2\right]}\biggr\}.
\eea
After calculating the derivatives one gets
\bea
D^{ij}(x,x') &=& -\frac{1}{\pi^2}\biggl\{\frac{2(\Delta{\bf x})^i(\Delta{\bf x})^j - \delta^{ij}\left[|\Delta{\bf x}|^2 + (\Delta t - i\epsilon)^2\right]}{\left[(t - t' - i\epsilon)^2 - |{\bf x} - {\bf x}'|^2\right]^3}
\nn\\
&+& \frac{2(\Delta{\bf x})^i(\Delta{\bf x})^j - \delta^{ij}\left[|\Delta{\bf x}|^2 + (\Delta t+i\epsilon)^2\right]}{\left[(t - t' + i\epsilon)^2 - |{\bf x} - {\bf x}'|^2\right]^3}\biggr\}.
\label{hada}
\eea
From the results above one may evaluate the Hadamard's elementary functions found in equation~(\ref{vfha3}). On the other hand, from equation~(\ref{pauli})  one can compute the Pauli-Jordan functions appearing in equation~(\ref{rrha3}).

\end{document}